# Flow characteristics and gelatinization kinetics of rice starch under strong alkali conditions

H. Yamamoto [*], E. Makita, Y. Oki and M. Otani

*Department of Food Science and Nutrition, Faculty of Human Life and Science, Doshisha Women's College of Liberal Arts, Kamigyo-Ku, Kyoto 602-0893, Japan*

**Abstract**

The normality dependence and the kinetics of flow behavior in the rice starch dispersion under strong alkali conditions were studied. Non-Newtonian flow (power-law) quantities $n$ and $\mu$ ($\sigma = \mu(\dot{\epsilon})^n$; $\sigma$: shear stress, $\dot{\epsilon}$: shear rate) of the samples stored (gelatinized) at 20°C for 10min after the addition of NaOH solution with the normality 0.090 - 0.175N, were measured by a cone-plate type rotational viscometer. For the flow analysis a care was taken for the ambiguity of the analysis range of flow data. The samples added with 0.14 - 0.16N NaOH displayed an apparently dilatant flow. As the normality of added NaOH increased, apparent viscosity $\eta_a$ was steeply enhanced in the vicinity of 0.14N and started to grow almost exponentially. Next, the storage time dependence of flow properties was investigated for samples gelatinized for $6 \leq t[\min] \leq 230$ after the addition of 0.146N NaOH solution. A Newtonian analysis of the flow ($\sigma$ - $\dot{\epsilon}$) data indicated that viscosity $\eta$ increased exponentially in the early stage ($6 \leq t \leq 35$) and did linearly in the middle. It was argued that the observed time dependence of viscosity could be described by a first-order reaction hypothesis combined with the fluidity ($\eta^{-1}$) mixing rule with respect to ungelatinized and gelatinized fractions of starch. The application of this theoretical treatment to the present data suggested that the gelatinization process was composed of plural phases with different values of a rate constant. For the observed kinetic characteristic of alkali gelatinization, the discussions were presented on the basis of amylose complex formation.

*Key words:* Rice Starch, Alkali Gelatinization, Viscosity, Kinetics, Mixing Rule

[*] Corresponding author. Tel.&Fax: +81-75-251-4215.
  *Email address:* `hyamamot@dwc.doshisha.ac.jp` (H. Yamamoto).

# Introduction

Starch gelatinization phenomena in a hot aquoes medium have long been one of principal subjects in food sciences from both applied and fundamental points of view. Although physicochemical fundamentals are still not fully understood, the influence of various factors, such as time, temperature and components (moisture, salts, sugar, lipids $\cdots$) on this phenomenon has been vigorously investigated. In recent years, the kinetic studies using differential scanning calorimetry (DSC) have become popular (Wirakartakusumah, 1981; Lund, 1984; Lund & Wirakartakusumah, 1984; Slade & Levine, 1991; Zanoni, Schiraldi, & Simonetta, 1995; Yeh & Li, 1996; Ojeda, Tolaba, & Suárez, 2000). The kinetic treatment is important in the study of starch gelatinization because the system during gelatinization is in a nonequilibrium state and the knowledges obtained from the kinetic studies could serve to clarify the reaction mechanism in molecular levels.

However, few attempts have so far been made to construct the starch gelatinization kinetics from the rheological viewpoint such as those based on flow properties of starch dispersions. This would partly be due to technical difficulties in carrying out rheological measurements with sample temperatures kept high and constant (65 - 90°C). In order to develop the simple kinetics of starch gelatinization, one must define a degree of gelatinization and follow its change as a function of time keeping all other environmental variables constant. In the DSC approach, a degree of gelatinization is directly measured from an enthalpy of gelatinization. On the other hand, in the rheological approach, one must define a well-defined physical (measureable) quantity which is related to a degree of gelatinization, and must measure its time dependence under isothermal conditions. The choice of an adequate physical quantity and its relation with a degree of gelatinization are not trivial.

Kubota, Hosokawa, Suzuki, and Hosaka (1979) studied the time dependence of fluid consistency index for the gelatinization of rice and potato starches using capillary viscometry. After gelatinizing the starch paste for a fixed time under isothermal conditions, they cooled it immediately to stop gelatinization and evacuated the gas. They then prepared the 3wt% starch dispersion and measured the flow behavior at 30°C. By assuming a first-order reaction model and defining the degree of gelatinization as a ratio of the fluid consistency index, they calculated the reaction rate constant.

In the present work, we focused on the flow properties of the starch dispersion gelatinized by adding sodium hydroxide (NaOH). Under strong alkali conditions starch is known to gelatinize even at room temperatures. Among the applications of this phenomenon is to utilize it for producing an adhesive paste. From an academic interest, this peculiarity of the alkali gelatinization could give us a technical advantage for the rheological measurement in isothermal conditions. Compared with thermal starch gelatinization, however, much less is known about alkali gelatinization (Suzuki & Juliano, 1975; Oosten, 1982, 1983, 1990; Otsuka, Moritaka, Fukuba, Kimura, & Ishihara , 2001; Moritaka, Ishihara, Matsumoto, Iino, & Kimura, 2003), and especially no kinetic approach has appeared.

Here we studied the flow properties of the dispersion of rice starch which is one of well-known seed-originated starches. Although some differences are present in some technical aspects of experimental and theoretical treatments, it would be interesting to compare a part of the present results with those of Kubota et al. (1979). Using a cone-plate type rotational viscometer, we first investigated the flow properties of the samples with a fixed storage (i.e. gelatinization) time after the addition of NaOH solution and their dependences on a normality of added NaOH. The results were analyzed by power-law model to elucidate the normality dependence of a flow behavior index and a consistency coefficient as well as an apparent viscosity. Especially we found that the samples prepared by NaOH with normalities in the vicinity of 0.15N exhibited nearly Newtonian flows.

Next, for the samples added with the NaOH solution of this fixed normality, for which the Newtonian approximation is expected to fit well in some time region, we studied the storage time dependence of their flow properties by using both power-law and Newtonian models. If the Newtonian model can be a good approximation model, we can define the normal viscosity as well as the apparent viscosity, as functions of time. These would enable us to consider the kinetics of starch gelatinization from the rheological viewpoint. For the obtained time dependence of viscosity, we found a novel concavity property of the viscostiy - time curve in the early stage of gelatinization. This kinetic characteristic of alkali gelatinization is qualitatively different from that known for thermal gelatinization. Inspired by that, we proposed a new type of the mixing rule of viscosity (fluidity) with respect to ungelatinized and gelatinized fractions of starch. It was argued that this choice of mixng rule could describe the time dependence of viscosity observed for the present alkali gelatinization of rice starch.

**Experimental**

*Sample preparation*

In this study the commercial powdered sample of nonglutinous rice starches "Better Friend", obtained from Shimada Chemical Co. (Japan), was used with-out any purification. The composition of the starch powder taken from supplier information is shown in Table 1.

The experimental samples for the study of the normality dependence were prepared in the following way. The weighed NaOH solutions (95.56g) at different levels of normality (0.090 - 0.175N) and the rice starch powder (4.44g) weighed into 100-m$\ell$ beakers, were stored at 20°C before mixing. At the equilibrium, the NaOH solution was added to the starch, at which the time measurement of the gelatinization period was to start ($t = 0$). The NaOH-starch mixture in the beaker was immediately stirred with a glass rod of 7-mm diameter by sufficient times to obtain a homogenious dispersion (actually 100 times). After stirring the beaker was stored in the incubator controlled at 20°C untill the flow measurement. For each normality, three samples were prepared in most cases.

The samples for studying the time dependence of viscosity were prepared in a similar manner as above except for the followings. The 0.146N NaOH solutions (20°C, 860.04g in all), divided into 4 portions, were added successively to the starch (20°C, 39.96g) weighed into a 1-$\ell$ beaker. The storage (gelatinization) time measuremnt was to start ($t = 0$) at which the NaOH solution was first added to the starch. Under the mixing, the mixture was stirred totally 900 times to avoid inhomogeneity. From this starch dispersion (900g), 8 portions of each 100g were poured into eight 100-m$\ell$ beakers. They were wrapped by polymethylpentene films to prevent drying and were stored in the incubator (20°C) until the flow measurement so that the total storage (gelatinization) time $t$[min] ranged from 6 to 230. 90sec before the flow measurement, each sample was stirred again 10 times by the glass rod.

The concentration of the starch powder in all samples thus prepared became 4.44wt%, and hence the substantial starch concentration was 3.8wt% if one considered only the carbonhydrate content of the starch (Table 1 supplied by the manufacturer).

*Flow measurement*

The flow measurement was made using a controlled stress rheometer CSL$^2$ 500 (TA Instruments) equipped with a cone (4cm diameter, 2° angle) and plate geometry. The cone-plate geometry provides uniform shear and is suited for studying a non-Newtonian fluid as well as a Newtonian one. The starch samples were poured onto the bottom plate controlled to be at 20°C with a Peltier system. Shear stress $\sigma$[Pa] was controlled to increase linearly from 0.0 to 2.0 in 2.0min. The time $t$[min] at which the flow measurement was made, i.e. the time at the end of the storage (gelatinization) period, was defined by just the middle time of measurement, in other words, 1.0min after the measurement had started. All the flow data of the present starch samples were collected and converted to the shear stress $\sigma$[Pa] - shear rate $\dot{\epsilon}$[s$^{-1}$] data by Rheology Solutions Software (TA Instruments, ver. 1.1.6).

*Flow analysis*

For studying the normality dependence, the samples were treated as non-Newtonian fluids and the converted flow data were analyzed by the power-law model

$$\sigma = \mu(\dot{\epsilon})^n \tag{1}$$

where $\mu$[Pa·s$^n$] and $n$ (dimensionless) denote respectively consisitency coefficient and flow behavior index. The flows are characterized in terms of these quantities as well as an apparent viscosity $\eta_a$ defined later.

As the result of power-law analysis we found that the samples prepared by NaOH in the vicinity of 0.15N showed nearly Newtonian flows ($n \approx 1$). We then performed measurements for the samples prepared by 0.146N NaOH to study the time dependence of flow properties to which the power-law analysis (1) was firstly applied. Next, we fixed flow behavior index $n$ to be 1, as a working hypothesis (an approximation). Namely, we analyzed the flow data by a Newtonian model

$$\sigma = \eta \dot{\epsilon} \tag{2}$$

with $\eta$[Pa·s] the viscosity of sample. The validity of this approximation should be checked statistically (in the sense of linear regression) and we shall actually find that this Newtonian approximation is fairy good ($R^2 > 0.90$) at least in the early stage of gelatinization. The viscosity $\eta$ defined by this Newtonian model and an apparent viscosity $\eta_a$ will be used for the kinetic arguments.

Let us explain our analysis procedures explicitly. For the Newtonian approximation, a linear regression analysis ($y = ax + b$) was applied to the obtained $\sigma$ - $\dot{\epsilon}$ data with $b$ fixed to be zero. Viscosity $\eta$ was obtained as the coefficient $a$. The power-law approximation model (1) is rewritten by

$$\log \sigma = \log \mu + n \log \dot{\epsilon} \tag{3}$$

The $\sigma$ - $\dot{\epsilon}$ data of each sample was hence converted to the $\log \sigma$ - $\log \dot{\epsilon}$ data to which a linear regression analysis was applied. From the results of the linear regression analysis then followed the values of flow behavior index $n$ and consisitency coefficient $\mu$. Although this procedure is not a complete non-linear regression analysis for the power-law approximation (1), we used this simple method as a working hypothesis in the present work. All this is the basic description of our flow analysis.

*Supplements for flow analysis methods*

As regards the details of flow analysis procedures, we need to give some technical supplements. With both use of flow analysis models (2) and (1), the results thus obtained should more or less depend on the choice of the analysis range of the flow data, i.e. the range of data points to which a regression analysis is applied. For example, the Newtonian approximation (2) is generally valid in a relatively smaller shear stress region. A deviation from the Newtonian law brought by the structure change of sample particles could occur in a large shear stress region. The result of the regression analysis would therefore strongly depend on the upper limit of the analysis (stress) range.

On the other hand, the power-law model usually continues to be an adequate analysis model to approximate the real flow curves of starch samples even under large shear stresses. A decrease of the apparent viscosity due to a large shear stress is reasonably reflected by a pseudo-plastic flow property ($n < 1$) of the power-law model. Rather, a linear regression performed for the converted data ($\log \sigma$ - $\log \dot\epsilon$) would be quite sensitive to unavoidable experimental fluctuations in a small shear stress region, and hence to the lower limit of the analysis range. It is simply because the logarithmic quantities $\log \sigma$ and $\log \dot\epsilon$ vary rapidly against the slight changes of $\sigma(\approx 0)$ and $\dot\epsilon(\approx 0)$ respectively.

To study the normality dependence, we made all the measurements at $t[\min] = 10$. For the flow analysis, considering the ambiguity as for the analysis range pointed above, we used the power-law model (1) with the following two kinds of analysis methods (procedures). One analysis method is to analyze the converted data ($\log \sigma$ - $\log \dot\epsilon$) in the whole range of measurement corresponding to the shear stress range $0 < \sigma[\mathrm{Pa}] < 2.0$ (the power-law analysis method I). By use of this method all informations included in the measurement were taken into account. The other analysis method is to analyze the converted data points within a variable $\log \sigma$ - $\log \dot\epsilon$ range corresponding to the shear stress range $y < \sigma[\mathrm{Pa}] < 2.0$, and to seek for the lower limit $y^*[\mathrm{Pa}]$ which could provide the highest regression $R^2$; the power-law model fits best in the range $y^* < \sigma[\mathrm{Pa}] < 2.0$ (the power-law analysis method II).

Based on the above measurement and the power-law analysis, we found that the samples prepared by NaOH with the normalities in the vicinity of 0.15N exhibited nearly Newtonian ($n \approx 1$) flows at $t[\text{min}] = 10$. Then, for the samples prepared with the 0.146N NaOH and stored (gelatinized) from $t[\text{min}] = 6$ to 230, we carried out flow measurements to see the storage (gelatinization) time dependence of their flow properties. To the obtained data we first perfomed the power-law analysis (1) with the above methods I and II, and investigated the time dependence of flow quantities $n$ and $\mu$.

After checking that the value of flow behavior index varies below and above $n = 1$ in some time region we next analyzed the flow data by taking $n$ to be a constant ($= 1$), i.e. by applying a Newtonian approximation (2) with the following two analysis methods (procedures). One method is to analyze the data points ($\sigma$ - $\dot{\epsilon}$) in the whole range of measurement $0 < \sigma[\text{Pa}] < 2.0$ (the Newtonian analysis method I). The other is to analyze the data points within a variable shear stress range $0 < \sigma[\text{Pa}] < y$ and to search for the upper limit $y^*[\text{Pa}]$ giving the highest validity $R^2$; the Newtonian model fits best in the range $0 < \sigma[\text{Pa}] < y^*$ (the Newtonian analysis method II). The validity of these Newtonian analyses should anyhow be checked by the $R^2$ values of linear regression analyses. We will actually see that the Newtonian approximation is fairy good ($R^2 > 0.90$) even for the method I at least in the early stage ($t[\text{min}] < 40$) of gelatinization where a novel kinetic property shall be found.

Now the final comment in this section is in order on the definition of apparent viscosity $\eta_a$ which shall appear in the following sections. The apparent viscosity in the present article means the ratio $\sigma/\dot{\epsilon}$ defined at the shear rate - shear stress data point ($\dot{\epsilon}, \sigma$) corresponding to the middle value ($\sigma[\text{Pa}] = 1.0$) of a whole shear stress range (0 - 2.0[Pa]). This data point was obtained at the middle time of each measurement, which was consistently the time $t[\text{min}]$ defined as the end point of a storage (gelatinization) time of the starch sample.

**Results**

*Normality dependence*

In Fig. 1 are exhibited the results of the normality (0.090 - 0.175N) dependence of flow behavior index $n$ and of consistency coefficient $\mu$ respectively. They were obtained by the measurement at $t[\text{min}] = 10$ after the addition of NaOH. For each quantity, solid square ■ and open square □ represent the data points derived by the power-law analysis methods I (whole stress region) and II (maximum $R^2$) respectively. In a normality region lower than 0.160N, the flow behavior index $n$ for the analysis method II showed higher values than those for the analysis method I.

The $R^2$ values for both analysis methods (Fig. 2) were considerably high ($\geq$ 0.9) except for a very high normality ($\geq$ 0.165N) region. For the power-law analysis method II, a ratio of the lower limit $y^*$[Pa] to the final value (2.0[Pa]) of the analysis range and its normality dependence are also displayed in Fig. 2. In a low normality ($\leq$ 0.14N) region all these quantities were stable against the normality change.

If we consider the flow behavior index $\bar{n}$ (not shown in the figure) averaged over those obtained by the power-law analysis methods I and II (Fig. 1), the whole normality region (0.09 - 0.175N) is divided into three regions by the value of $\bar{n}$. The fow behavior index $\bar{n}$ is larger than 1 (dilatancy) in the normality region 0.14 - 0.16N and is smaller (pseudo-plasticity) in other two regions. We note that the presence of the dilatancy in a certain normality region (0.15 - 0.16N) was also checked by use of the power-law analysis (the whole stress region) computer program which is proper to the flow mode of Rheology Solutions Software (TA Instruments, ver. 1.1.6) although the detailed statistical procedure for its regression analysis is not manifest in this software.

Consistency coefficient $\mu$ (Fig. 1) seems to grow abruptly at some normalities. For non-Newtonian fluids these changes of strength in a viscous property could be quantitatively described in terms of apparent viscosity. Fig. 3 displays the logarithm of apparent viscosity $\log_{10} \eta_a$ and its dependence on the normality of added NaOH. This revealed that apparent viscosity $\eta_a$[Pa · s] staying at $10^{-3}$ - $10^{-2}$ scales in the region 0.09 - 0.14N, was abruptly enhanced when the normality goes beyond 0.14N. After that, apparent viscosity seems to grow exponentially with the increase of normality (0.14 - 0.17N), and at 0.175N reaches almost $10^2$[Pa · s] scales.

*Time dependence: power-law analysis*

Fig. 4 shows flow behavior index $n$ and consistency coefficient $\mu$ respectively as functions of a storage time ($6 \leq t$[min] $\leq 230$) after the addition of 0.146N NaOH solution. In the time region $10 \leq t$[min] $\leq 23$, the flow behavior index $n$ obtained by the power-law analysis method II(□) was larger than 1, whereas the index $n$ obtained by the method I (■) was smaller than 1. In that region, the $R^2$ value for the method I was reduced to approximately 0.9 (Fig. 5), and the shear stress range obtained by the analysis method II (Fig. 5) was roughly $1 < \sigma$[Pa] (i.e. $y^*$[Pa] $\approx 1$) where the power-law approximation fitted best (maximum $R^2$).

These results indicated that the flow behavior of the sample is qualitatively different between in a smaller half ($\sigma$[Pa] $< 1$) and in a larger half ($1 < \sigma$[Pa] $< 2$) of the whole stress region. The typical $\sigma$ - $\dot{\epsilon}$ curve showing such a behavior is displayed in Fig. 6a where the data at $t[\min] = 13$ were plotted in the whole stress region and a solid line (a Newtonian approximation) was drawn as a reference.

From $t[\min] = 23$ to 40, however, such a discrepancy for flow behavior index $n$ was considerably reduced (Fig. 4). This fact was also observed for the $R^2$ value (Fig. 5) and for the analysis range obtained by the method II (Fig. 5). The typical result ($t[\min] = 25$) is depicted in Fig. 6a with the Newtonian approximation (a solid line). The power-law analysis method I provided $n = 0.91$ with a markedly high fitting ($R^2 = 0.991$).

After $t[\min] = 100$, the $R^2$ value for the power-law analysis method I was reduced to being smaller than 0.90 (Fig. 5). However, it was always kept high ($R^2 > 0.95$) for the power-law analysis method II. These results indicated that after $t[\min] = 100$ the flow behavior involved irregularity in a smaller stress region and that the power-law model (1) did not fit consistently in the whole stress region. The typical example is exhibited in Fig. 6b ($t[\min] = 170$) with a power-law approximation (solid) curve, for which the power-law analysis method I gives $n = 0.63$ with an extremely low fitting $R^2 = 0.515$ (Fig. 5).

If we judge from the time dependence of the flow behavior index $\bar{n}$ (data not shown) averaged over those obtained by the power-law analysis methods I and II (Fig. 4), the sample showed an apparently dilatant flow ($n > 1$) from $t[\min] = 12$ to 21. On the other hand the time dependence of consistency coefficient $\mu$ (Fig. 4) suggested that the viscosity growth of the sample might be accelerated in some restricted time region ($t[\min] < 50$). Such a kinetic characteristic, however, must be verified with other well-defined quantities because flow behavior index $n$ and hence the dimension of consistency coefficient $\mu$ varied radically in the region $t[\min] < 50$ (Fig. 4); one could not make a quantitative evaluation about the difference of consistency coefficient $\mu(t_1) - \mu(t_2)$ unless $n(t_1) = n(t_2)$.

In order to legally discuss about the increasing rate of some physical quantity that could characterize the change of flow properties, namely to construct the kinetics of starch gelatinization in the language of rheology, we must consider a quantity with a constant dimension, such as the viscosity defined by the Newtonian analysis.

*Time dependence: Newtonian analysis*

In Fig. 7 are exhibited the storage time dependences of viscosities obtained by using the Newtonian analysis methods I and II, and of the apparent viscosity $\eta_a$. Although some disorder was generated in the data distributions at $t[\min] > 50$, the three $\eta$ - $t$ curves were regular and almost identical to each other (Fig. 8) in the region $t[\min] < 40$ where the validity of the Newtonian analysis was sufficiently high ($R^2 > 0.90$) even in the whole stress region. Further, as shown in Fig. 8, the viscosity - time curve possesses a globally downward curvature

$$\frac{d^2\eta}{dt^2} > 0 \qquad (4)$$

This concave property of the $\eta$ - $t$ curve in the early stage of alkali gelatinization was not observed in the case of the corresponding thermal gelatinization (Kubota et al, 1979), and might suggest an essential difference between the kinetic processes of these two types of starch gelatinizations.

Actually, a careful analysis of the $\eta$ - $t$ curves clarified that for each definition of viscosity an exponential fitting proved to be a good approximation to a certain region of each $\eta$ - $t$ curve. In Fig. 8 are displayed the best exponential fittings (solid curves) to the viscosity - time curves obtained respectively by the Newtonian analysis method I ($6 \le t[\min] \le 35$, $R^2 = 0.993$), II ($6 \le t[\min] \le 36$, $R^2 = 0.995$) and as apparent viscosity ($6 \le t[\min] \le 35$, $R^2 = 0.992$). In this figure linear approximations (dotted lines) were also added to demonstrate the concave property of the $\eta$ - $t$ curves.

After the exponential growth region, viscosity seemed to rise almost linearly in the middle region $t[\min] \ge 36$ (Fig. 7). This fact was quantitatively checked for all definitions of viscosity although the best linear regions were not common among them.

The value of viscosity took the maximum roughly at 160 - 170[min] (Fig. 7). The gelatinized starch might then reach an equilibrium state, but the viscosity - time data were not stable. As seen from Figs. 4, 5 and 6b the $\sigma$ - $\dot{\epsilon}$ data themselves were irregular in that time region, and a Newtonian property was lost (Fig. 4). This irregularity of the $\eta$ - $t$ data might be a disadvantage of the present experimental institution, i.e. a rather high starch concentration and the measurement by a cone-plate type viscometer. If the regular data could be obtained by some adequate method, the $\eta$ - $t$ data should be observed to approach smoothly to the equilibrium value $\eta_G$ with the convex property $d^2\eta/dt^2 < 0$.

On the other hand, the $\sigma$ - $\dot{\epsilon}$ data were regular (see e.g. Fig. 6a) at 10 - 40[min], and a quite high validity ($R^2 > 0.93$) was achieved for the Newtonian approximation to them. Further, the $\eta$ - $t$ data had very little dependence on the analysis method (Fig. 8). We therefore conclude that viscosity of rice starch dispersion increased exponentially in the early stage (10 - 36[min]) of alkali gelatinization. This result is different even qualitatively from that known for the ordinary case, i.e. the thermal gelatinization of rice starch in a neutral pure water (Kubota et al, 1979).

**Theoretical analysis**

The experimental results for the time dependence of viscosity presented in the previous section suggested the necessity to introduce a new idea for the kinetic treatment of the starch gelatinization under strong alkali conditions. In this section we propose the new treatment based on the general idea of the mixing rule of viscosity for ungelatinized and gelatinized fractions of starch under gelatinization, and compare the derived consequence with the experimental results.

*Starch gelatinization kinetics with mixing rule*

Starch gelatinization phenomenon is described as a kind of chemical reaction process, in which an ungelatinized part UG of starch is changed to a gelatinized part G. Then, the sum of their amounts [UG] and [G] should equal to the amount of an ungelatinized part UG at the initial time (denoted by the symbol 0) of gelatinization, i.e.

$$[\text{UG}] + [\text{G}] = [\text{UG}]_0 \tag{5}$$

The fundamental hypothesis of gelatinazation kinetics is that the ungelatinized part obeys a general $n$-th order reaction rate equation

$$-\frac{d[\text{UG}]}{dt} = K_n [\text{UG}]^n \tag{6}$$

where $K_n$ denotes an $n$-th order reaction rate constant. Gelatinization degree $x$ and hence ungelatinization degree $1-x$ are defined respectively as the ratios of [G] and [UG] to [UG$_0$]

$$x \equiv \frac{[\text{G}]}{[\text{UG}]_0}, \qquad 1 - x = \frac{[\text{UG}]}{[\text{UG}]_0} \tag{7}$$

where the second equality follows from (5). Using this relation one can rewrite a general $n$-th order reaction rate equaion (6) as

$$\frac{dx}{dt} = \tilde{K}_n (1-x)^n \tag{8}$$

where $\tilde{K}_n \equiv K_n [\text{UG}]_0^{n-1}$ is a new $n$-th order reaction rate constant of (time)$^{-1}$ dimension. In most of the past researches, people have made sure from various approaches that the simplest choice $n = 1$ worked for starch gelatinization kinetics (e.g. Lund (1984)). Under the initial condition $x(t = 0) = 0$, equation (8) with $n = 1$, i.e. the first-order reaction rate equation possesses the solution

$$x = 1 - e^{-K_1 t} \tag{9}$$

Many investigators have so far related the above gelatinization degree $x$ with physically, chemically and biochemically measured quantities (e.g. Lund (1984)). Among them, Kubota et al (1979) expressed gelatinization degree $x$ in terms of the viscosity or exactly speaking the fluid consistency index $K$ in their literature [1] by

$$x = \frac{K - K_0}{K_e - K_0} \tag{10}$$

where $K_0$ and $K_e$ denote respectively the fluid consistency indices at the initial time and at the equilibrium. Replacing these consistency indices $K$, $K_0$ and $K_e$ in (10) by the corresponding viscosities $\eta$, $\eta_{UG}$ and $\eta_G$, we get the relation

$$\eta = \eta_{UG}(1 - x) + \eta_G x \tag{11}$$

with $\eta$ the viscosity of the sample dispersion under gelatinization. Here $\eta_{UG}$ and $\eta_G$ denote respectively the proper viscosities of ungelatinized (UG) and gelatinized (G) parts of the sample satisfying $\eta_{UG} < \eta_G$.

Equation (11) can be regarded as an additive mixing rule of viscosity with respect to ungelatinized and gelatinized fractions. Substituing of the first-order solution (9) for $x$ in the mixing rule (11) leads to the storage time dependence of viscosity

$$\eta(t) = \eta_G - e^{-K_1 t}(\eta_G - \eta_{UG}) \tag{12}$$

Its second derivative with respect to time $t$ is calculated with the result

$$\frac{d^2\eta}{dt^2} = -(\eta_G - \eta_{UG})K_1^2 e^{-K_1 t} < 0 \tag{13}$$

implying that the viscosity - time curve is always convex at $t < \infty$. This consequence had really held in the case of thermal gelatinization (Kubota et al, 1979) but is clearly in contradiction to the present experimental results. If we want to keep the first-order reaction hypothesis (9), we are then forced to seek for other types of mixing rule than the formula (11).

---

[1] Their definition of the fluid consistency index $K$ is different from that of the consistency coefficient $\mu$ used here. The flow behavior index in their literature corresponds to the inverse of that used in the present article. Also, one should not confuse it with our symbol $K_n$ of a reaction rate constant.

The mixing rule (11) can be regarded as the special case ($\nu = 1$) of the following general power-law formula

$$\eta^\nu = \eta_{\text{UG}}^\nu (1 - x) + \eta_{\text{G}}^\nu x \tag{14}$$

with an exponent $\nu$ a free dimensionless mixing parameter. We consider this free parameter to reflect the various characteristics of sample starch dispersions such as starch spieces and/or environmental conditions.

Then, another typical case of (14) could be that with $\nu = -1$, in which instead of viscosity $\eta$ itself fluidity, i.e. the inverse of viscosity $\eta^{-1}$ should obey the additive law

$$\phi = \phi_{\text{UG}}(1 - x) + \phi_{\text{G}} x \tag{15}$$

where $\phi(\equiv \eta^{-1})$ represents the fluidity of the sample dispersion, and $\phi_{\text{UG}}(\equiv \eta_{\text{UG}}^{-1})$ and $\phi_{\text{G}}(\equiv \eta_{\text{G}}^{-1})$ should satisfy the condition $\phi_{\text{UG}} > \phi_{\text{G}}$. Substituing of the first-order solution (9) for (15) provides the fluidity as a function of a storage time, expressed by

$$\phi(t) = \eta^{-1}(t) = \phi_{\text{G}} + e^{-K_1 t}(\phi_{\text{UG}} - \phi_{\text{G}}) \tag{16}$$

From this the second derivative of viscosity is calculated with the result

$$\frac{d^2 \eta}{dt^2} = \frac{K_1^2 (\phi_{\text{UG}} - \phi_{\text{G}}) e^{-K_1 t}}{(\phi_{\text{G}} + (\phi_{\text{UG}} - \phi_{\text{G}})) e^{-K_1 t})^3} \left( (\phi_{\text{UG}} - \phi_{\text{G}}) e^{-K_1 t} - \phi_{\text{G}} \right) \tag{17}$$

Since the first fractional factor in the right-hand side is always positive, the sign of (17) is determined by the second factor $D(t) \stackrel{\text{def}}{=} (\phi_{\text{UG}} - \phi_{\text{G}}) e^{-K_1 t} - \phi_{\text{G}}$, which is a monotonously decreasing function of gelatinization time $t$, and is negative at the equilibrium $D(\infty) = -\phi_{\text{G}}$. Therefore, if $D(0) = \phi_{\text{UG}} - 2\phi_{\text{G}}$ is positive, i.e. $\eta_{\text{G}} > 2\eta_{\text{UG}}$, the sign of second derivative $d^2\eta/dt^2$ becomes positive before a certain gelatinization time $t^*$

$$\frac{d^2 \eta}{dt^2} > 0 \quad (0 \leq t < t^*) \tag{18}$$

Note that the condition $\eta_{\text{G}} > 2\eta_{\text{UG}}$ could reasonably be satisfied for real starch samples including the present case. The theoretical result (18) implies that the shape of the viscosity - time curve is *concave* in the first half ($t < t^*$) of gelatinization process.

*Application of the theory ($\nu = -1$) to the experimental results*

The logical consequence derived from the above kinetic treatment based on the combination of the first-order solution (9) with the mixing rule of fluidity (15) seems qualitatively consistent with the storage time dependence of viscosity observed for the present alkali gelatinization of rice starch. Irrespective of the definitions of viscosity, the viscosity - time curve was actually concave in the early ($t[\min] < 40$) stage of the gelatinization process (Fig. 8). This concave property of the $\eta$ - $t$ curve was not seen in the corresponding $K$ - $t$ curve for the thermal gelatinization of rice starch (Kubota et al, 1979), and cannot be theoretically understood by combining the first-order reaction hypothesis with the mixing rule of viscosity (11). In this subsection, we would like to give some more quantitaive comparison with the experimental data.

Under the condition $\eta_{\mathrm{UG}} \ll \eta_{\mathrm{G}}$, which is actually possible, equation (16) can be legitimately reduced to

$$\phi(t) = \eta^{-1}(t) = \phi_{\mathrm{UG}} e^{-K_1 t} \tag{19}$$

unless an inequality $K_1 t \gg 1$ is satisfied. This means that viscosity should exhibit an exponential increase

$$\eta(t) = \eta_{\mathrm{UG}} e^{K_1 t} \tag{20}$$

in a certain initial stage of gelatinization. As demonstrated in Fig. 8 we in fact confirmed this exponential increase to fit with the observed viscosity - time data in the region $6 \leq t[\min] \leq 35$ or 36. The fitting was attained with markedly high accuracy ($R^2 > 0.99$) for all definitions of viscosity. From these exponential analyses (Fig. 8), a rate constant $K_1[\min^{-1}]$ was estimated to be 0.165 - 0.168, and the value of $\eta_{\mathrm{UG}}[\mathrm{Pa \cdot s}]$ was predicted to be about $1.2 \times 10^{-3}$. This scale of $\eta_{\mathrm{UG}}$ is smaller by $\mathrm{O}(10^{-4})$ compared with that of $\eta_{\mathrm{G}}$ evaluated from the experimental result (Fig. 7), and proved the consistency ($\eta_{\mathrm{UG}} \ll \eta_{\mathrm{G}}$) of the present exponential reduction.

Without using such a truncation of the theory, however, we can make a quantitative comparison between the full content (16) of the theory and the experimental data. Equation (16) can be exactly transformed to

$$\ln(\phi - \phi_{\mathrm{G}}) = -K_1 t + \ln(\phi_{\mathrm{UG}} - \phi_{\mathrm{G}}) \tag{21}$$

This relation states that the quantity $\ln(\phi - \phi_G)$ converted from the experimental $\eta$ - $t$ data should be linear in $t$ with the coefficient $K_1$. Then, the precise estimation of a rate constant $K_1$ only requires to know from the experiment a constant $\phi_G(= \eta_G^{-1})$.

As we have already remarked in the previous section, the present experimental data involved some irregular distribution at the final stage of gelatinization. Although the viscosity at the equilibrium $\eta_G$ was therefore not definite (Fig. 7), we shall take, as a working hypothesis, the values of $\eta_G$[Pa·s] to be 30 for the Newtonian analysis method I, and 35 for the method II and for apparent viscosity. These values were roughly estimated from the data in Fig. 7.

For each definition of viscosity, the quantity $\ln(\phi - \phi_G)$ was then plotted as a function of storage time. In Fig. 9 is shown the result for the viscosity obtained by the Newtonian analysis method I. We have checked that for all definitions of viscosity a high linearity ($R^2 > 0.90$) was maintained in a rather broad data region $6 \leq t[\min] \leq 110$. The value of a rate constant $K_1[\min^{-1}]$ was about 0.10 - 0.11, which would give an averaged rate constant of gelatinization in almost all of the total reaction time.

A further investigation revealed that the above time region was divided into plural regions with different values of a rate constant $K_1$. For example, the $\ln(\phi - \phi_G)$ - $t$ data of the Newtonian analysis method I was separated into at least three regions (Fig. 9).

First, in the region $6 \leq t[\min] \leq 35$ the highest linearity ($R^2 = 0.993$) was realized with $K_1[\min^{-1}] = 0.166$. Secondly, within the remaining range $36 \leq t$, the $R^2$ value became highest (= 0.967) in the region $36 \leq t[\min] \leq 55$ with the smaller value of a rate constant $K_1[\min^{-1}] = 0.103$. In the third region $56 \leq t[\min] \leq 130$, $R^2$ took the highest value ($R^2 = 0.896$) in the rest ($56 \leq t$) of the time range, and a rate constant was further reduced to be $K_1[\min^{-1}] = 0.042$.

The $\ln(\phi - \phi_G)$ - $t$ data derived by the Newtonian analysis method II and those from apparent viscosity indicated similar behaviors to the above, although these data are not displayed here. Especially the rate constant $K_1[\min^{-1}]$ in the first region $6 \leq t[\min] \leq 35$ or 36 took almost the same values 0.167($\pm$0.01), which were also very close to those obtained by an exponential approximation (Fig. 8). They were all reduced in the later stages.

**Discussions**

*Difference of mixing rule*

As seen in the previous section, the viscosity of rice starch under the present alkali gelatinization showed the kinetic behavior which is different from that of the corresponding thermal gelatinization (Kubota et al., 1979). This difference was mathematically characterized by different forms of mixing rule (11) and (15), in other words, the different relations between viscosity and gelatinization degree.

In order to gain an insight into the physical meaning of these mixing rules, it is helpful to describe the gelatinization process controlled by these typical mixing rules in terms of the flow composition models (Fig. 10). In this model, the flow part $\boxed{\longrightarrow}$ with a long arrow expresses UG with a large proper fluidity $\phi_{\text{UG}}$ (a small viscosity $\eta_{\text{UG}}$) and that $\boxed{\rightarrow}$ with a short arrow does G with a small proper fluidity $\phi_{\text{G}}$ (a large viscosity $\eta_{\text{G}}$).

Although the initial ($x = 0$) and final ($x = 1$) states are common, mixed states in the middle (e.g. $x = 1/2$) of reaction are different between $\nu = 1$ and $\nu = -1$ (Fig. 10). The additive viscosity mixing rule with $\nu = 1$ expressed in terms of fluidity

$$\phi^{-1} = \phi_{\text{UG}}^{-1}(1 - x) + \phi_{\text{G}}^{-1}x \tag{22}$$

is symbolized by the series-type composition model of fluidity where shear stress is additive and shear rate is kept constant in the mixed state. The viscosity of whole sample increases linearly with the increase (decrease) of the high (low) viscous fluid $\boxed{\rightarrow}$ ($\boxed{\longrightarrow}$).

In contrast, the additive mixing rule of fluidity ($\nu = -1$) given by equation (15) is symbolized by the parallel-type composition model of fluidity where shear stress is kept constant and shear rate is additive in the mixed state. This intuitively suggests that in the initial stage of gelatinization the fluidity of the whole sample is largely dominated by UG $\boxed{\longrightarrow}$ and that the increase of G does not directly (lineary) affect to the increase of the viscosity of whole sample.

To make this intuitive argument more transparent, we introduce the dimensionless fluidities $\Phi \equiv \phi/\phi_{\text{UG}}(= \eta_{\text{UG}}/\eta)$ and $\Phi_{\text{G}} \equiv \phi_{\text{G}}/\phi_{\text{UG}}(= \eta_{\text{UG}}/\eta_{\text{G}} < 1)$ and rewrite the mixing rules (22) and (15) as

$$\Phi^{-1} = 1 + (\Phi_G^{-1} - 1)x \qquad (\nu = 1) \tag{23}$$
$$\Phi = 1 - (1 - \Phi_G)x \qquad (\nu = -1) \tag{24}$$

For the latter ($\nu = -1$) the derivatives of (dimensionless) viscosity $\Phi^{-1}$ with respect to gelatinization degree are calculated with the results

$$\frac{d\Phi^{-1}}{dx} = \frac{1 - \Phi_G}{(1 - (1-\Phi_G)x)^2}, \quad \frac{d^2\Phi^{-1}}{dx^2} = \frac{2(1-\Phi_G)^2}{(1-(1-\Phi_G)x)^3} > 0 \tag{25}$$

and therefore

$$\left.\frac{d\Phi^{-1}}{dx}\right|_{x=0} = 1 - \Phi_G < \Phi_G^{-1} - 1 = \left.\frac{d\Phi^{-1}}{dx}\right|_{\nu=1} \tag{26}$$

This implies that in the initial stage of gelatinization, the viscosity increasing ratio to gelatinization degree is smaller in the case $\nu = -1$ than that in the case $\nu = 1$. In Fig. 11 is described the example of whole relative viscosity $\Phi^{-1}$ - gelatinization degree $x$ curves controlled by both mixing rules ($\nu = \pm 1$) where $\Phi_G$ is commonly set to be 0.1. In both cases the dimensionless viscosity $\Phi^{-1}$ starts from 1 and increases toward $\Phi_G^{-1}$ ($= 10$ in Fig. 11), but along different paths.

For $\nu = 1$, the whole viscosity $\Phi^{-1}$ increases lineary with the increase (decrease) of G (UG). Whereas for $\nu = -1$, even if G occupies partially ($x \ll 1$) UG dominates the whole fluidity, as was intuitively suggested from the flow composition model (Fig. 10). Hence the viscosity increase with $\nu = -1$ is milder than that with $\nu = 1$, and the relative viscosity $\Phi^{-1}$ - gelatinization degree curve becomes concave (Fig. 11), which contributes to the concavity property of the first half ($t < t^*$) of the viscosity - time curve (18). This would physically indicate that in the initial stage ($x \ll 1$) of the gelatinization the independecy of G is so high that its appearance does not directly contribute to the viscosity increase of whole sample (UG + G). The result of our experiment showed that this propery is realized for the kinetic characteristic of alkali gelatinization of rice starch.

*Hypothesis of alkali induced complex formation*

Then, we should next discuss the physicochemical reason why such a rheologically high independency of G is realized only for alkali gelatinization. Physically starch consists of amorphous and crystallite regions, and the gelatinization initially occurs in the former. It was proposed that the swelling of amorphous phase contributes to the disruption of the crystallites and that the gelatinization is a (semi-)cooperative process (Marchant & Blanshard, 1978;

French, 1984). On the other hand, starch is chemically a mixed system of amylose and amylopectin, and for the starches from cereal grains such as rice starch (the present sample) the former forms an inclusion complex with lipids. The major fatty acids of lipids bonded to rice starch are palmitic and linolenic acids.

This amylose - lipid complex originally present in the ungelatinized starch is known to suppress the gelatinization and swelling of starch granules, although this complex would become unstable and might break down at very high temperatures. In the present experiment, the gelatinization proceeded at 20°C and hence the amylose - lipid complexes would not be instabilized thermodynamically.

However, the added NaOH could bring a chemical instability to original amylose - lipid complex. I t is widely recognized that amylose is able to capture many kinds of compounds including alkali hydroxides such as NaOH and to form various types of complexes (see Tomasik & Schilling (1998a,b) for recent comprehensive reviews). Such complex formations of amylose could generally affect the gelatinization behavior, and in some cases were reported to bring drastic changes (Lindqvist, 1979; Larsson, 1980).

In addition to the direct complex formation between amylose and NaOH, another possibility of a new strong complex formation of amylose caused by NaOH addition must be pointed out. After NaOH penetrates into starch granule, the lipids originally included in amylose could react with NaOH to generate the sodium salts of higher fatty acids. These sodium salts (soap) of higher fatty acids such as palmitic acid, should work as strong surfactants with large hydrophobic groups and hydrophilic groups, and could form new strong and stable complexes with amylose. The main driving force of this complex formation is a hydrophobic bond, which is further strengthened by the existence of a hydrophilic group with polarity. Some of the ionic surfactants were suggested to be able to complex even with amylopectin (Evans, 1986).

Actually such an effect of surfactants or emulsifiers to form strong complexes with starches, has been utilized in modifying the texture of starch products (Krog, 1971). Gray and Schoch (1962) reported that polar surfactants which complex strongly with amylose restricted the swelling and solubilization of various starches over the pasting range 60 - 95°C.

At the present low experimental temperatures (20°C) the amylose - fatty acid complexes are expected to be thermodynamically stable. These strong and stable amylose complexes, thus generated in the amorphous region, would form isolated objects in starch granule. As the result, in comparison with thermal gelatinization, the initial stage of alkali gelatinization would become unable to cooperatively induce strain in the crystallites of amylopectin; the swelling speed of starch granule is thus suppressed in the initial stage. This situation might be linked with the rheologically high independency of G and the reduction of the influence on UG. Anyhow we hypothesize that such a structural change of the gelatinized state of starch could lead to the observed kinetic difference of viscosity. To elucidate the detailed mechanism, however, further studies would be necessary.

## Conclusions and remarks

### Conclusions

The flow characteristics of rice starch (3.8wt%) gelatinized by NaOH solution (20°C) were studied. The power-law and the Newtonian models were adopted in analyzing the measured flow data. Viscosity was defined in three kinds of manners according to the analysis procedures.

When 0.14 - 0.16N NaOH were added, the starch dispersion stored for 10[min] showed an apparently dilatant flow. As the normality of NaOH increased, apparent viscosity was rapidly enhanced in the vicinity of 0.16N and started to grow almost exponentially.

The storage (gelatinization) time dependence of viscosity was investigated for the sample added with 0.146N NaOH, and turned out to be qualitatively different from that of thermal gelatinization in a neutral pure water. Remarkably, the viscosity - time curve showed an exponential growth in the early stage of gelatinization. This fact was confirmed irrespective of different definitions of viscosity, and was imcompatible with the first-order kinetic treatment which was previously taken for thermal gelatinization.

We thus proposed a kinetic treatment involving a viscosity mixing rule of general power-law type. It was argued that the time dependence of viscosity in the present alkali gelatinization could be described by a first-order reaction hypothesis combined with the mixing rule of fluidity. The application of this theoretical treatment to the experimental data suggested that the alkali gelatinization process of rice starch dispersion consisted of plural phases characterized by different values of a rate constant.

### Remarks

As the device of viscosity measurement we have used the cone-plate type rotational viscometer . This instrument was adequate to our purpose since it was useful not only for the Newtonian analysis but also for the power-law analysis to the flow data. For the time dependence of viscosity, we have checked that at least qualitatively the similar results to the present ones were obtained for dilute rice starch dispersions by performing a continuous measurement with use of a capillary viscometer. The continuous capillary viscometry is appropriate only for the measurement of apparent viscosity of dilute dispersions but could provide regular flow data even after a long gelatinization time.

It is not definite that the (apparent) dilatancy property observed in the samples added with 0.14 - 0.16N NaOH is really an intrinsic property of starch dispersion, because the present sample under measurement is not in the equilibrium state. The gelatinization of sample and hence the viscosity growth should slightly proceed even in 2.0 min of measurement. Such a "time effect" cannot be avoided in the present experimental situation and would always work toward enlarging flow behavior index $n$, i.e. dilatancy. On this point, further investigations would be necessary to have a definite conclusion.

In the theoretical aspect we have restricted the premise of our kinetic treatment to the first-order ($n = 1$) reaction model and to the typical choices of a mixing parameter $\nu = \pm 1$. Some possible arguments including a general $n$-th order kinetics with a general value of a mixing parameter $\nu$, as well as the experimental results obtained by the capillary viscometry, will appear in forthcoming articles. The influences of temperature, starch species and other solutes than NaOH would also be interesting subjects to be pursued experimentally in the future.

## Acknowledgements

We are indebted to Ms. S. Fujii, Ms. H. Kato, Ms. M. Nagahata, Ms. K. Saka and Ms. H. Shinada for their cooperations in the early stage of this work. One of the authors (H.Y.) would like to thank Ms. C. Shimizu for her help in analyzing the flow data.

## Figure Captions

Fig. 1. The flow behavior index $n$ and the consistency coefficient $\mu$ of the rice starch dispersion measured at $t[\text{min}] = 10$ after the addition of NaOH solution, and their dependences on the normality (0.090 - 0.175N) of added NaOH. The data series represented in terms of solid square ■ and open square □ are obtained by the power-law analysis methods I (whole stress region) and II (maximum $R^2$) respectively.

Fig. 2. The normality (0.090 - 0.175N) dependences on the $R^2$ values for the power-law analysis methods I (■) and II (□), and on the ratio of the lower limit $y^*[\text{Pa}]$ to the final value (2.0[Pa]) of the shear stress range obtained for the power-law analysis method II.

Fig. 3. The logarithm ($\log_{10}$) of apparent viscosity $\eta_a[\text{Pa·s}]$ at $\sigma[\text{Pa}] = 1.0$ of the rice starch dispersion measured at $t[\text{min}] = 10$ after the addition of NaOH solution, and its dependence on the normality (0.090 - 0.164N) of added NaOH. Apparent viscosity is defined by the data point of shear stress - shear rate curve at the middle value (1.0[Pa]) of the whole shear stress range (0 - 2.0[Pa]) applied to sample.

Fig. 4. The flow behavior index $n$ and the consistency coefficient $\mu$ of the rice starch dispersion stored for $6 \leq t[\text{min}] \leq 230$ after the addition of 0.146N NaOH solution. The data series represented in terms of solid square ■ and open square □ were obtaind by the power-law analysis methods I (whole stress region) and II (maximum $R^2$) respectively.

Fig. 5. The storage time $t[\text{min}]$ dependences on the $R^2$ values for the power-law analysis methods I (■) and II (□), and on the ratio of the lower limit $y^*[\text{Pa}]$ to the final value (2.0[Pa]) of the shear stress range obtained for the power-law analysis method II (maximum $R^2$).

Fig. 6. The shear stress - shear rate data at $t[\text{min}] = 13$ and 25 (a) with Newtonian approximations (solid lines) in the whole stress region, and that at $t[\text{min}] = 170$ (b) with a power-law approximation (solid curve) in the whole stress region.

Fig. 7. The viscosity [Pa·s] of the rice starch dispersion stored for $6 \leq t[\text{min}] \leq 230$ after the addition of 0.146N NaOH solution. The data series represented in terms of solid square ■, open square □ and solid triangle ▲ were obtaind respectively by the Newtonian analysis methods I (whole stress region), II (maximum $R^2$) and as an apparent viscosity $\eta_a$ at $\sigma[\text{Pa}] = 1.0$.

Fig. 8. The viscosity [Pa·s] of the rice starch dispersion stored for $t[\text{min}] = 6$ to 35 (a,c) and 36 (b) after the addition of 0.146N NaOH solution. The data series represented in terms of solid square ■ (a), open square □ (b) and solid triangle ▲ (c) were obtained respectively by the Newtonian analysis methods I, II and as an apparent viscosity. Solid curves are exponential fittings to the data series and dotted lines depict linear approximations.

Fig. 9. The quantity $\ln(\phi - \phi_\text{G})$ obtaind by the Newtonian analysis method I (whole stress region) for the rice starch dispersion stored for $6 \leq t[\text{min}] \leq 130$ after the addition of 0.146N NaOH solution. Three lines are linear approximations to the data in the time regions $6 \leq t \leq 35$, $36 \leq t \leq 55$ and $56 \leq t \leq 130$ respectively. The value of $\eta_\text{G}$ ($\equiv \phi_\text{G}^{-1}$) was chosen to be $30[\text{Pa·s}]$.

Fig. 10. The flow composition models describing the mixing rules $\nu = \pm 1$. The flow parts $\boxed{\longrightarrow}$ with a long arrow and $\boxed{\rightarrow}$ with a short arrow represent respectively the ungelatinized part with the proper fluidity $\phi_\text{UG}$ and the gelatinized part with the proper fluidity $\phi_\text{G} (< \phi_\text{UG})$.

Fig. 11. The relative viscosity $\Phi^{-1}$ - gelatinization degree $x$ curves controlled by the power-law mixing rules $\nu = \pm 1$. The relative proper viscosity $\Phi_\text{G}^{-1}$ is set to be 10 in both cases.

Table 1
The composition of rice starch powder, informed from the manufacturer, Shimada Chemical Co. (Japan).

| Ingredients | wt% |
|---|---|
| Carbohydrate | 85.3 |
| Moisture | 13.5 |
| Lipid | 0.7 |
| Protein | 0.3 |
| Ash | 0.2 |
| Ca | 0.029 |
| Na | 0.011 |

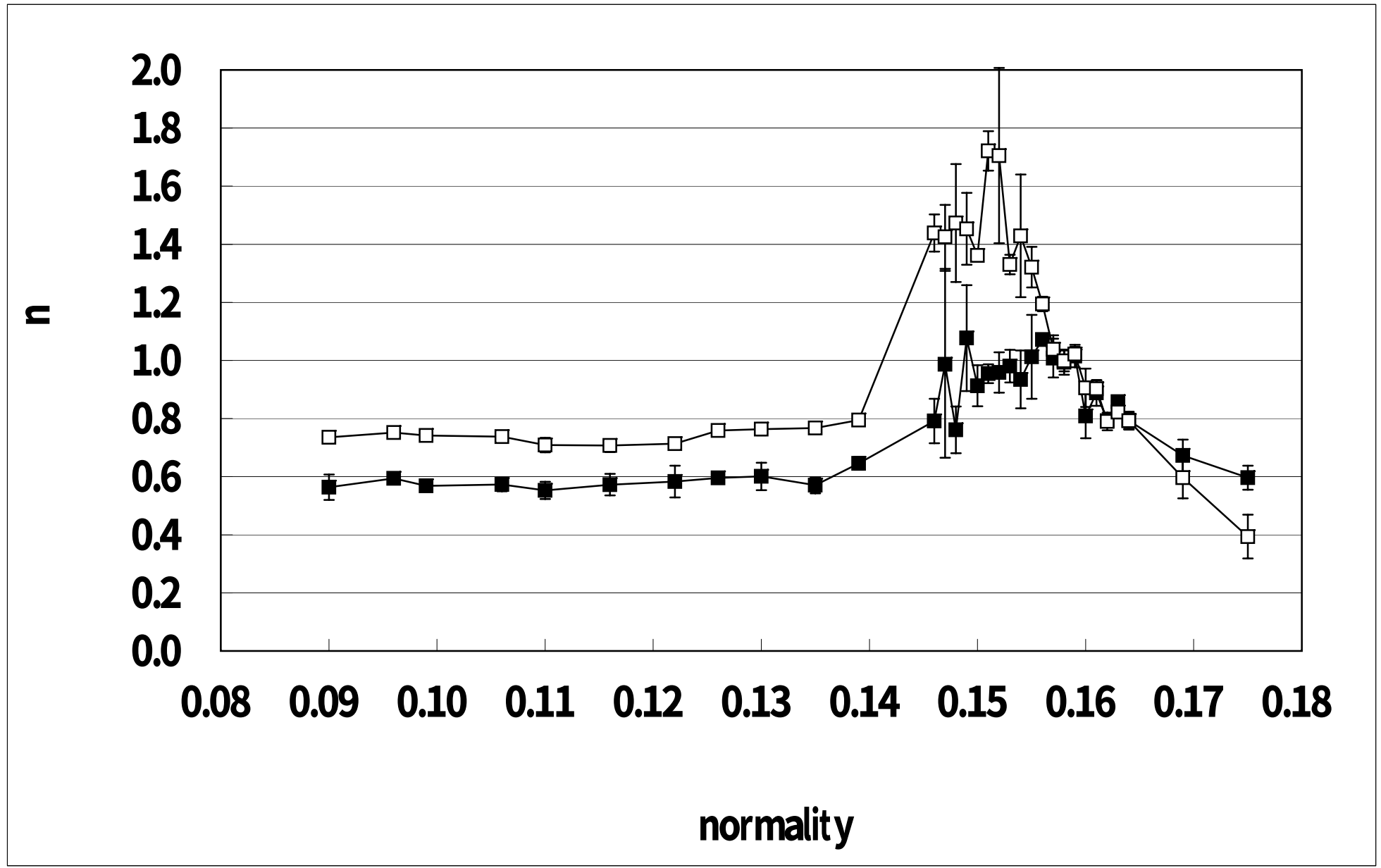

Fig 1

Hisashi Yamamoto

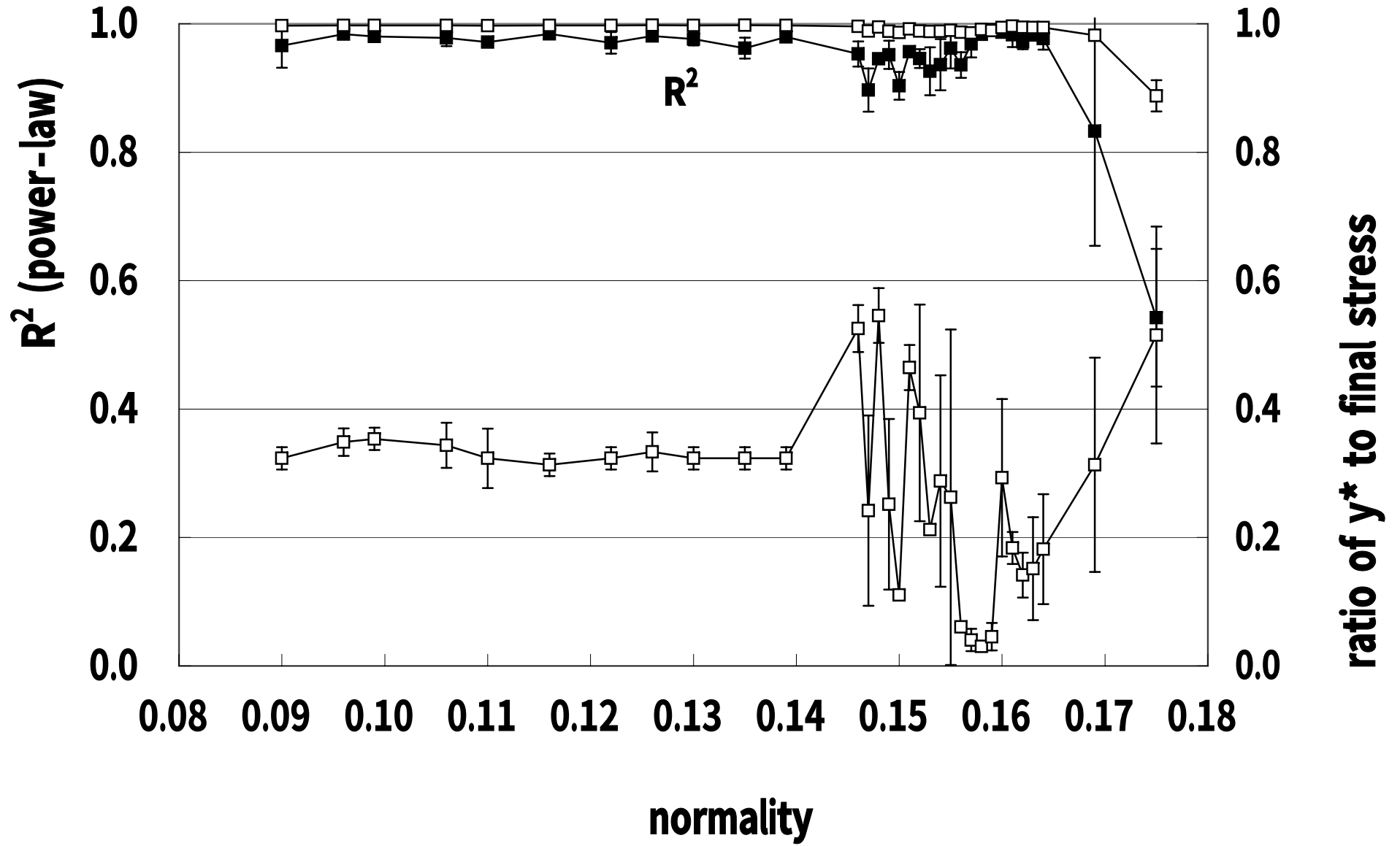

Fig. 2　Hisashi Yamamoto et al.

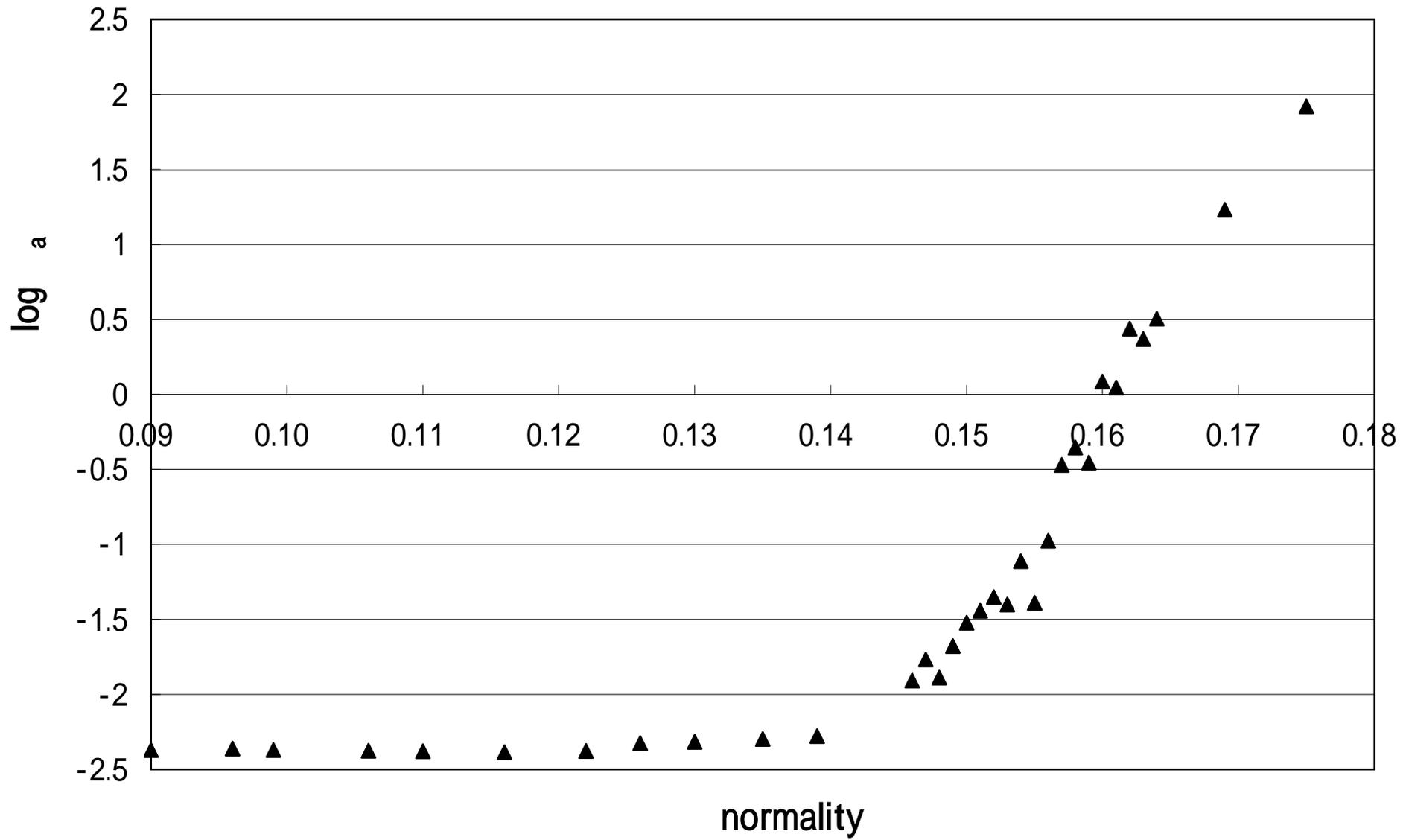

Fig. 3

Hisashi Yamamoto et al.

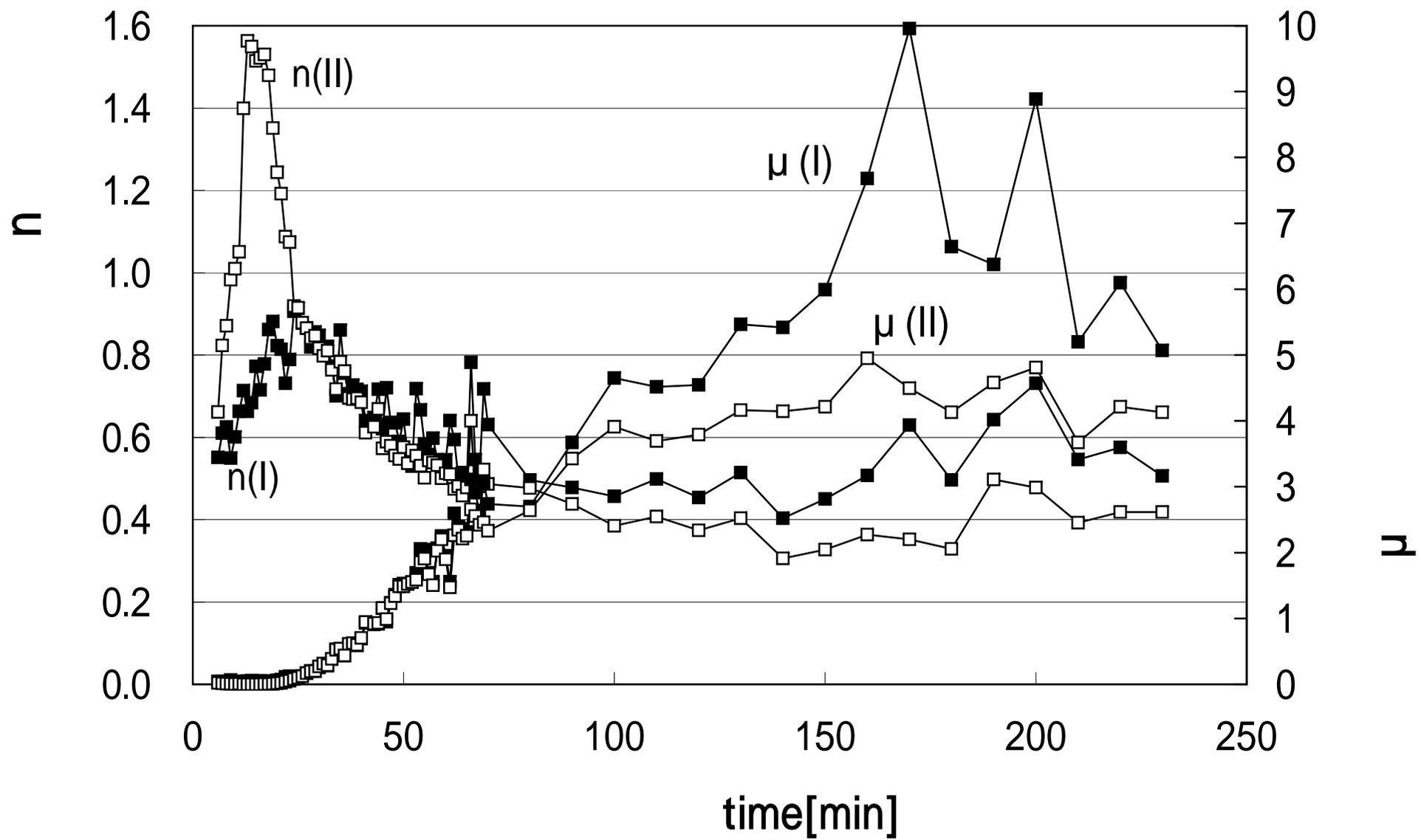

Fig. 4   Hisashi Yamamoto et al.

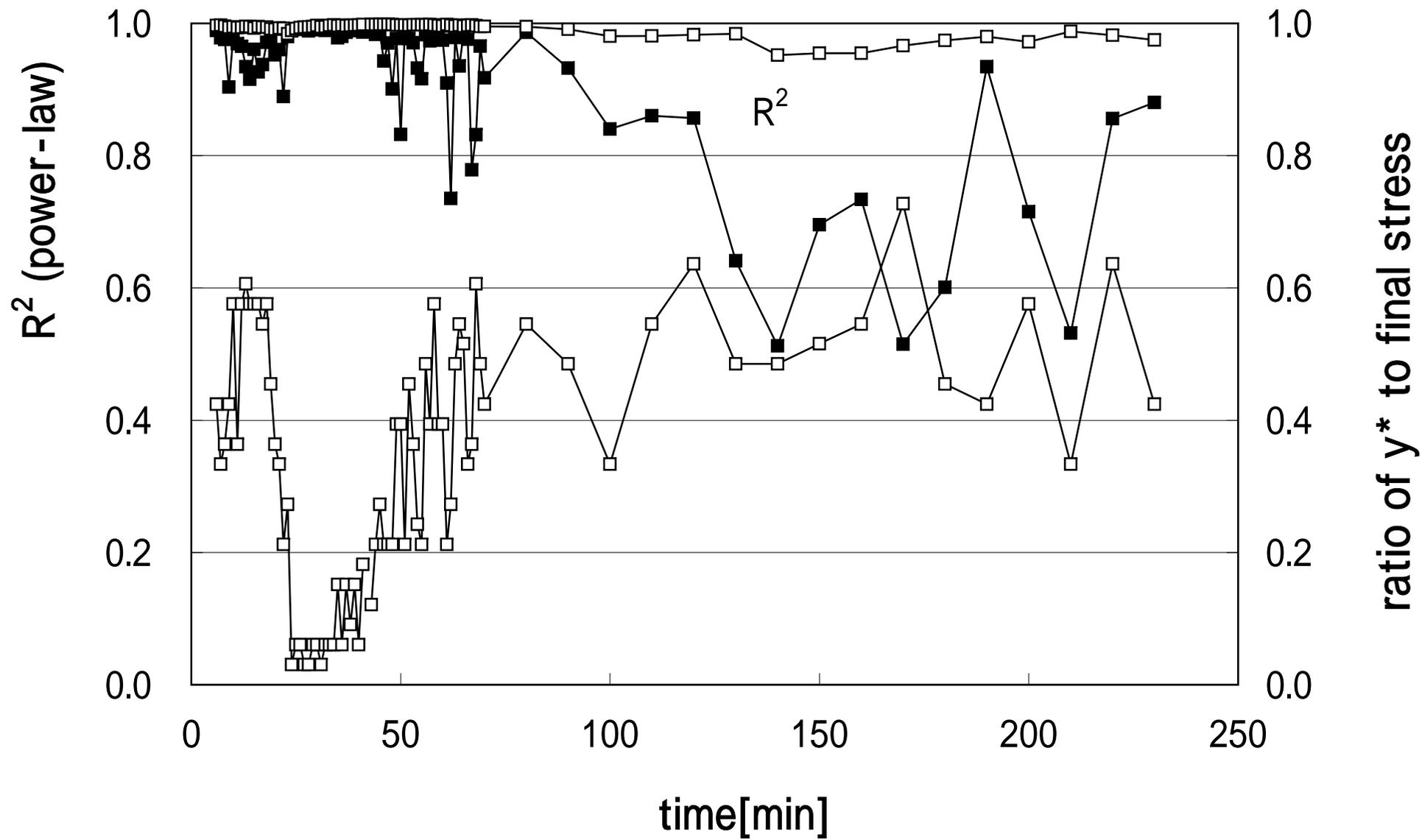

Fig. 5                                                                                                   Hisashi Yamamoto et al.

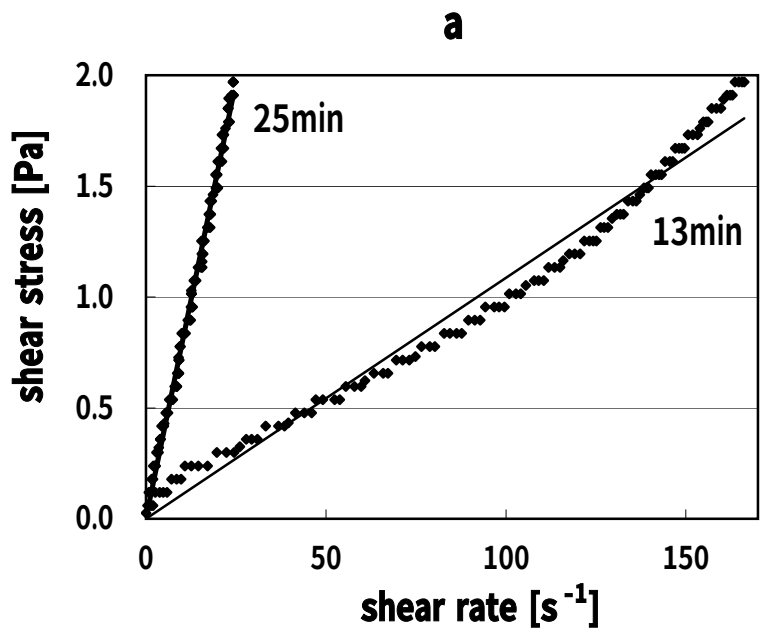 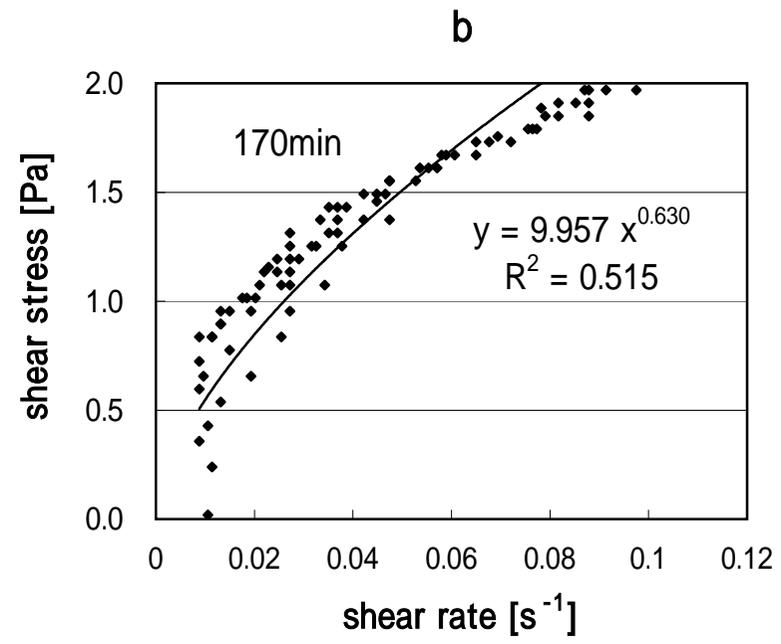

Fig. 6 Hisashi Yamamoto et al.

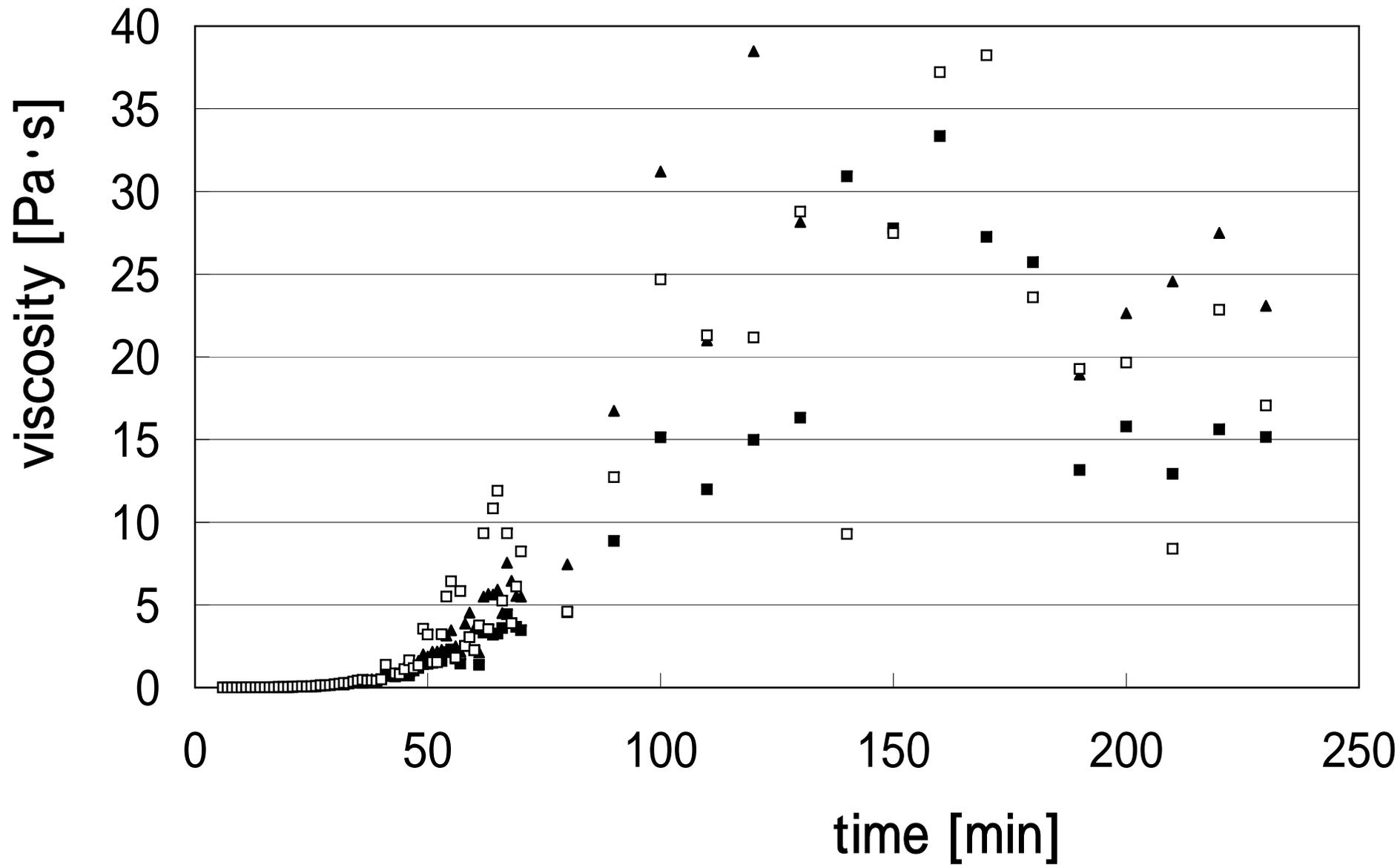

Fig. 7

Hisashi Yamamoto et al.

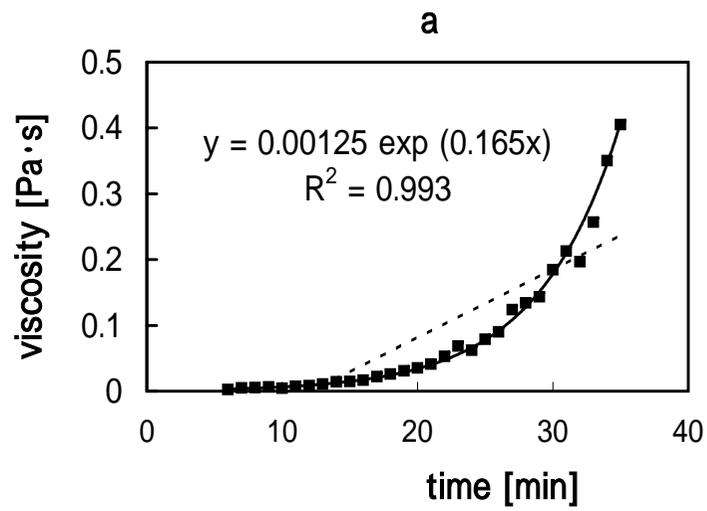
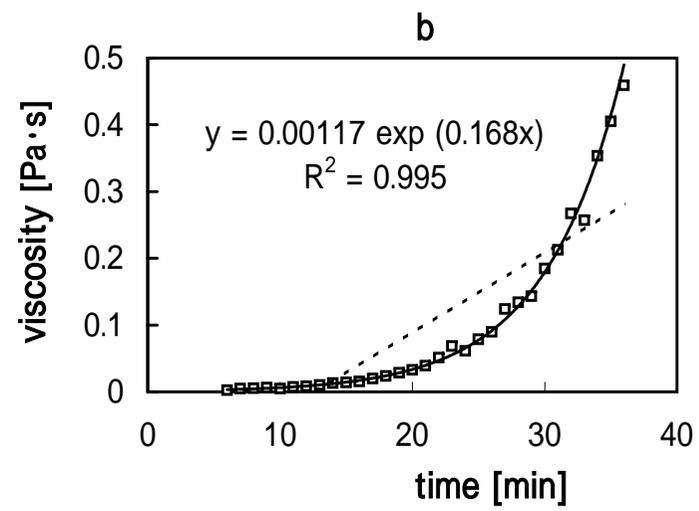
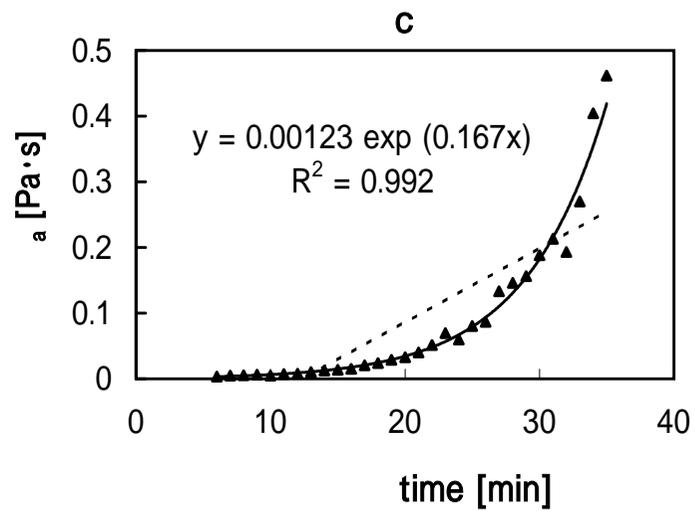

Fig. 8  Hisashi Yamamoto et al.

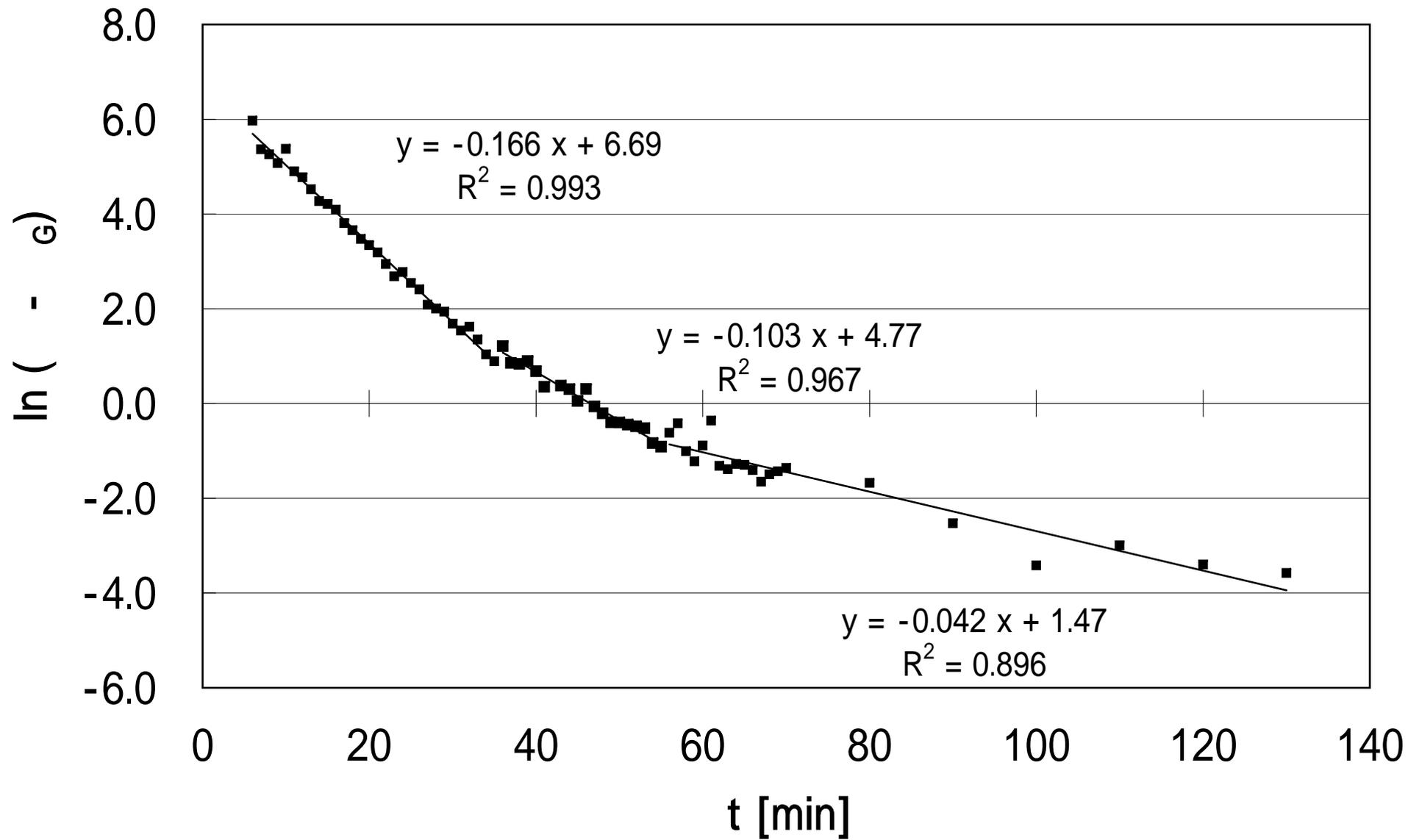

Fig. 9

Hisashi Yamamoto et al.

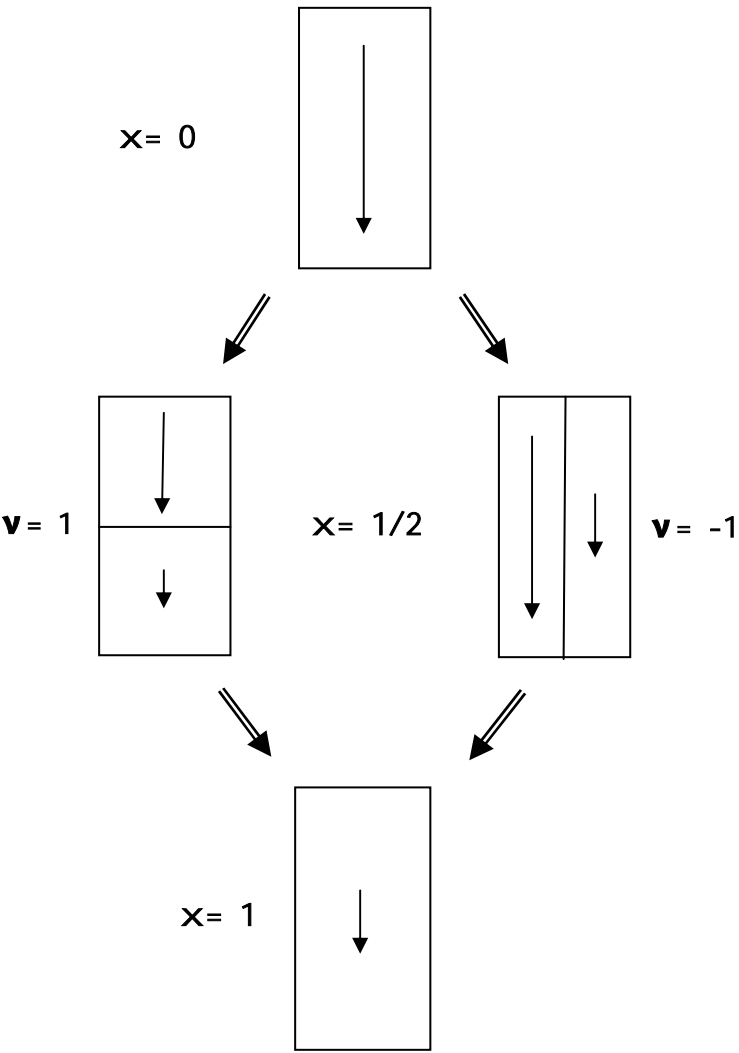

Fig. 10                      Hisashi Yamamoto et al.

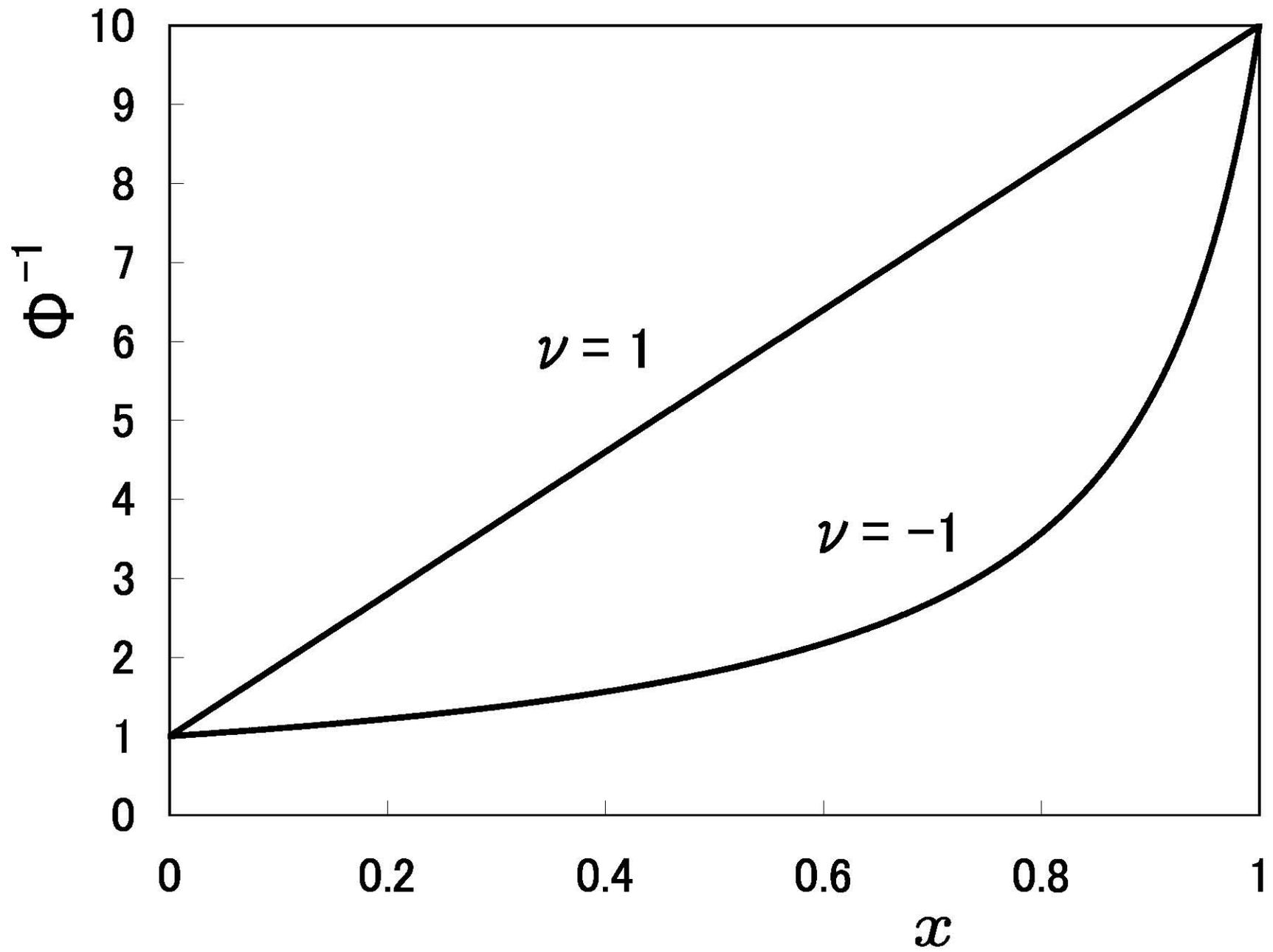

Fig. 11

Hisashi Yamamoto et al.